\documentclass[aps, prl,twocolumn, 10pt, showpacs, longbibliography, superscriptaddress]{revtex4-1}
\usepackage{graphicx}
\usepackage{dcolumn}
\usepackage{bm}
\usepackage{color}
\usepackage{amsmath, amsthm, amssymb}
\usepackage[modulo]{lineno}
\usepackage[usenames,dvipsnames]{xcolor}
\usepackage{ulem}
\newcommand{\ba}{\begin{eqnarray}}
\newcommand{\ea}{\end{eqnarray}}

\usepackage{float}
\floatstyle{boxed}

\begin{document}
\title{Development of high frequency and wide bandwidth Johnson noise thermometry}
\author{Jesse Crossno}
\affiliation{Department of Physics, Harvard University, Cambridge, MA 02138}
\author{Xiaomeng Liu}
\affiliation{Department of Physics, Harvard University, Cambridge, MA 02138}
\author{Thomas A. Ohki}
\affiliation{Raytheon BBN Technologies, Quantum Information Processing Group, Cambridge,
Massachusetts 02138, USA}
\author{Philip Kim}
\affiliation{Department of Physics, Harvard University, Cambridge, MA 02138}
\author{Kin Chung Fong}
\affiliation{Raytheon BBN Technologies, Quantum Information Processing Group, Cambridge,
Massachusetts 02138, USA}
\date{\today}
\begin{abstract}%
{We develop a high frequency, wide bandwidth radiometer operating at room temperature, which augments the traditional technique of Johnson noise thermometry for nanoscale thermal transport studies. Employing low noise amplifiers and an analog multiplier operating at 2~GHz, auto- and cross-correlated Johnson noise measurements are performed in the temperature range of 3 to 300~K, achieving a sensitivity of 5.5~mK (110 ppm) in 1 second of integration time. This setup allows us to measure the thermal conductance of a boron nitride encapsulated monolayer graphene device over a wide temperature range. Our data shows a high power law (T$^{\sim4}$) deviation from the Wiedemann-Franz law above T$\sim$100~K.}

\end{abstract}
\maketitle 
As modern electronics continue to miniaturize and increase in operation frequency, heat dissipation has begun to bottleneck their performance~\cite{Pop:2010p147}. Understanding thermal transport in these complex devices requires new thermometry techniques capable of dealing with the challenges unique to nanoscale systems~\cite{Cahill:2014p011305}. Diminishing system sizes call for measurements to be ¬¬¬non-invasive to avoid thermal agitation of minute heat capacities. As weak electron-phonon coupling can result in different electronic and lattice temperatures~\cite{Sergeev:2005p2209, Fong:2013p2542}, separate probes must be used to measure each temperature independently. Various mesoscopic experimental techniques, such as resistive thermometry, normal-insulator-superconductor thermometry, Coulomb blockade thermometry, and shot noise thermometry~\cite{Pobell, Giazotto:2006p1978, Spietz:2003p2080}, have been developed to meet these requirements. However, a versatile thermometry that works in a wide range of temperatures and experimental conditions (such as under strong magnetic field) has yet to be developed at the nanoscale.

Fundamentally based upon the fluctuation-dissipation theorem~\cite{Kubo:1966p2745}, Johnson noise thermometry (JNT) is a primary thermometry having a straight forward interpretation, independent of the material details, that stands out as a natural candidate for nanoscale thermal measurements. Temperature is measured by passively monitoring fluctuations of the conducting components within the nanoscale device without current excitations. Difficulties in measuring low level voltage noise~\cite{Pobell} has limited the applications of JNT to metrology~\cite{White:1996p2532, Benz:2011p2837} and extreme environments such as nuclear reactors~\cite{Brixy:1971p2839} and ultralow temperatures~\cite{Pobell}. Recently, considerable progress has been made in the measurement of noise on mesoscopic devices by applying radiometry techniques~\cite{ Spietz:2003p2080, Fong:2012p2256, Fong:2013p2542}, opening up the possibility of using JNT to study a wide range of mesoscopic phenomena. In this letter, we incorporate room temperature low noise amplifiers and an analog linear multiplier operating at 2~GHz to develop auto- and cross-correlated JNT for nanoscale devices. The higher operation frequencies allow wider bandwidth measurements which improves temperature resolution while the room temperature amplifiers allow JNT in strong magnetic fields. A precision of 0.01\% is achieved on auto- and cross-correlated noise measurements in 1 second of integration time. We demonstrate the capability of this setup by measuring the electronic thermal conductance---defined as the inverse of thermal resistance---of a mesoscopic graphene device over a temperature range of 3-300~K.

Johnson noise results from the spontaneous thermal fluctuation of charges in dissipative electrical elements at finite temperature. The time averaged mean-square voltage, $\langle V^2\rangle$, across a resistance, $R$, is described by the Nyquist theorem, $\langle V^2\rangle = 4Rk_BT_e\Delta f$ where $k_B$ is the Boltzmann constant, $T_e$ is the electron temperature, and $\Delta f$ is the equivalent noise bandwidth of the system. The Nyquist theorem holds in the limit of the photon energy being much smaller than the thermal energy, such that $hf \ll k_B T_e$, where $h$ is the Planck constant and $f$ is the frequency. The full quantum mechanical description is given by~\cite{Callen:1951p2747}
\begin{equation}
\label{JNeqFull}
\langle V^2\rangle =4hf~Re(Z)\left[\frac{1}{2}+\frac{1}{\exp(hf/k_BT)-1}\right] \Delta f
\end{equation}
where $Z$ is the complex impedance of the dissipative element. The first term in Eqn.~(1) is the zero point motion of the photon field while the second term is the one-dimensional blackbody radiation that our radiometer detects~\cite{Dicke:1946p2064}. The spectrum is frequency independent until rolling off above $k_BT_e/h$. At 1~K, this roll off is centered at 20~GHz, which sets the upper bound of the JNT operating frequency. 

In typical mesoscale conducting samples, a characteristic value of the channel resistance is on the order of $h/e^2 \sim$25~k$\Omega$, much larger than the 50~$\Omega$ characteristic impedance typical of high frequency, low noise amplifiers (LNA). To account for this mismatch, the Nyquist theorem can be rewritten to describe the Johnson noise power absorbed by an amplifier with characteristic impedance $Z_0$, as $\langle P\rangle = k_B T \Delta f [1-(\frac{Z-Z_0}{Z+Z_0})^2]$. For large sample impedance, an LC tank circuit can be employed to couple the noise power into the amplifier ~\cite{Schoelkopf:1998p2255, Fong:2012p2256}. 
\begin{figure}[t]
\includegraphics[width=1\columnwidth]{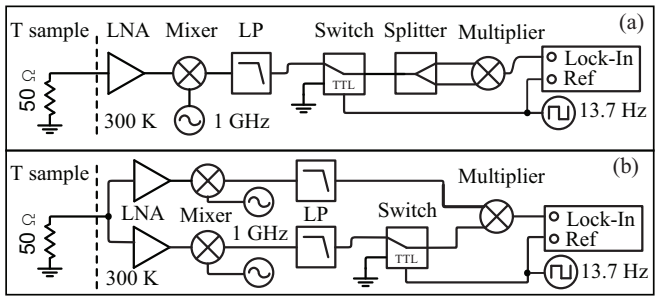}
\caption{Simplified electrical schematic for the auto- (a) and cross- correlation (b) Johnson noise thermometry setups. The low noise amplifiers (LNA) have a noise temperature of $\sim$50~K. Bandwidth is defined by a homodyne mixer and low-pass filter. Linear multiplication is performed at frequencies up to 2~GHz by an RF multiplier that acts as a square-law detector (a) or a cross-correlator (b). A microwave switch chops the signal at 13.7~Hz, modulating the signal away from dc.}
\label{fig:ExpSchematic}
\end{figure}

Fig.~1(a) shows the schematic of our auto-correlation JNT setup. The resistive load is connected directly to the LNA for Johnson noise amplification. The signal-to-noise ratio of the noise measurement is mostly determined by this front-end LNA~\cite{Pozar:MicrowaveBook}. The SiGe LNA (Caltech CITLF3) used in this report has a room temperature noise figure, in the frequency range of 0.01 to 2~GHz, of about 0.64~dB, corresponding to a noise temperature of 46~K. A low pass filter and homodyne mixer define a bandwidth of 328~MHz centered at 1~GHz. The center frequency should be high enough to avoid $1/f$ fluctuations and allow wide bandwidth noise measurements, while low enough to avoid stray capacitance in the device and the high frequency Johnson noise roll off. An analog linear multiplier and RF power splitter detects the noise power. Operating from dc to 2~GHz, the multiplier, Analog Devices ADL5931, serves as a square law detector with 30~dB dynamic range. The noise power is modulated by a microwave switch at 13.7~Hz and subsequently measured by a lock-in amplifier.
\begin{figure}[t]
\includegraphics[width=1\columnwidth]{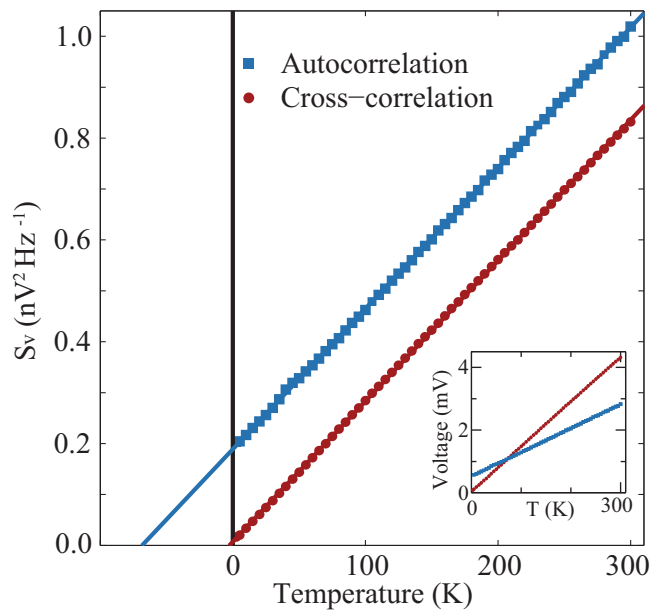}
\caption{Johnson noise of a 50~$\Omega$ resistive load as measured by auto- and cross- correlation setups, Fig. 1 (a) and (b), respectively. The inset shows the lock-in amplifier outputs as a function of bath temperature. These signals are converted to noise power in the main panel using the Nyquist equation. The solid lines are linear fits, where the auto- and cross- correlation data exhibit an offset of 68~K and 2.6~K, respectively, due to amplifier noise.}
\label{fig:JohnsonNoise}
\end{figure}

By varying the temperature $T$ of the resistive load attached to a cold finger in a cryostat, the auto-correlated Johnson noise data is collected. As shown in the inset of Fig.~2, the signal is linear in device temperature and offset by the constant amplifier noise over the measured temperature range of 3-300~K. Calibration is done through the Nyquist equation resulting in the main panel of Fig.~2. The autocorrelation data is best described by the equation: $S_V = \mathcal{G}k_B(T_e+T_{N})$ where $\mathcal{G}$ is the proportional gain factor set by the LNA amplification together with the insertion loss of microwave components and $T_{N}$ is the total system noise temperature. As $T_e$ tends to zero, Johnson noise subsides but the system noise remains. We find that our auto-correlation setup has a $T_{N}$ of 68~K, consistent with the LNA specification at room temperature. We note that this can be further reduced by lowering the LNA operating temperature.

Dissipation between the resistive load and the LNA, such as coaxial attenuation and contact resistance, can contaminate thermal transport measurements~\cite{White:1996p2532, Glattli:1997p7350}. Johnson noise from the sample is added to the unwanted Johnson noise from these lossy components. Cross-correlation techniques can mitigate this problem by amplifying the Johnson noise signal of interest independently via two separate measurement lines~\cite{Glattli:1997p7350, DiCarlo:2006p2620, Henny:1999p2790, Brophy:1965p2623, *Klein:1978p2616} and discarding uncorrelated noise between the two channels. Previously, cross-correlation measurements were limited to frequencies below a few MHz due to the practical implementation of multipliers and digital processing speeds~\cite{Brophy:1965p2623, *Klein:1978p2616, Sampietro:2000p2679, DiCarlo:2006p2620}. The 2~GHz analog multiplier and LNA, combined with the lock-in amplifier modulation scheme described in Fig.~1(b), measure the correlated noise between the two channels, rejecting a large portion of the uncorrelated amplifier noise. Residual cross talk~\cite{Glattli:1997p7350, Sampietro:2000p2679} between the two channels offsets the data by 2.6~K. 
\begin{figure}[t]
\includegraphics[width=1\columnwidth]{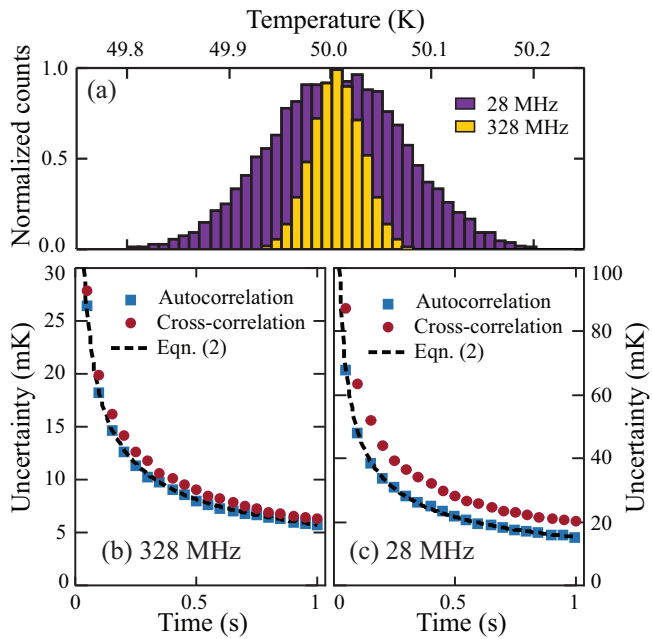}
\caption{(a) Histograms of 20,000 auto-correlation temperature measurements for 28 and 328~MHz bandwidth using 50~ms integration time. Histogram peaks are normalized to 1 for clarity. Standard deviation of 1000 temperature measurements as a function of integration time for (b) 328~MHz and (c) 28~MHz bandwidth. All data is taken on a 50~K resistive load with uncertainties approaching 100 ppm.}
\label{fig:Sensitivity}
\end{figure}

Johnson noise is a stochastic process with statistical uncertainties determined by the integration time and measurement bandwidth. The noise of our Johnson noise thermometer can be characterized by repeating a measurement multiple times and studying how it fluctuates about the mean. Fig. 3(a) compares two histograms, both containing 20,000 autocorrelation measurements at 50~K with 50~ms integration time but using two different bandwidths: 28 and 328~MHz. Wider bandwidth reduces the statistical uncertainties in Johnson noise thermometry, leading to the better performance demonstrated at 328~MHz. Fig. 3(b-c) plots the standard deviation of 1000 temperature measurements for 28~MHz and 328~MHz correlation bandwidth, respectively, as a function of integration time. In both auto- and cross-correlation experiments, the sensitivity follows an inverse square root power law described by the Rice relation~\cite{Rice:1944p2823, White:1996p2532}:
\begin{equation}
\label{Rice}
\delta T=\frac{T_e+T_{sys}}{\sqrt{\Delta f \cdot\tau}}
\end{equation}
where $\delta T$ is the measurement uncertainty, $\tau$ is the integration time and $T_{sys}$ characterizes the statistical noise in the system. While cross-correlation reduces the noise offset shown in Fig. 2 compared with autocorrelation, their statistical fluctuations are comparable as $T_{sys}$ is dominated by the amplifier noise. We attain 5.5~mK uncertainties on a 50~K signal (0.01\% precision) in 1 second of integration time.
\begin{figure}[t]
\includegraphics[width=1\columnwidth]{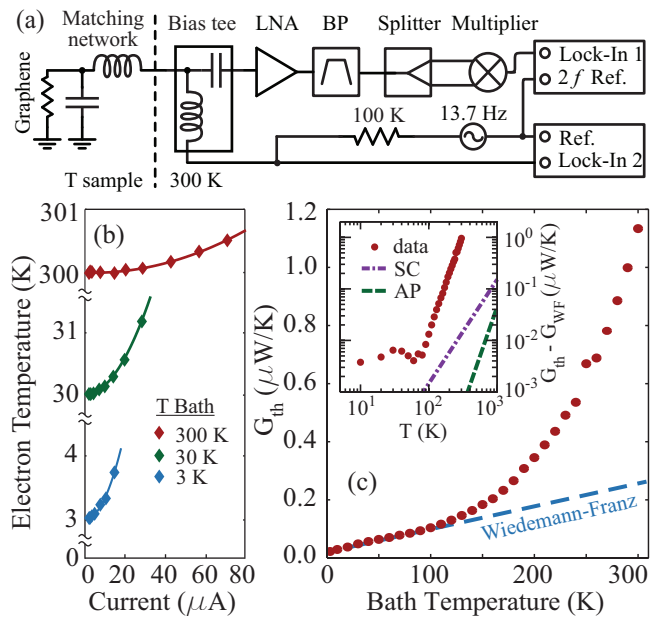}
\caption{(a) Schematic of thermal conductance measurement. Graphene is impedance matched to $\sim$50~$\Omega$ in the measurement bandwidth by an on chip, LC tank circuit. Low frequency heating current is injected through the bias tee while high frequency Johnson noise is monitored similar to Fig.~1(a). (b) Electron temperature rise in graphene under Joule heating. Solid lines are quadratic fits with $G_{th}$ as the only fitting parameter. (c) Graphene thermal conductance from 3~K to 300~K. Blue dashed line shows the theoretical Wiedemann-Franz conductance for our device geometry offset by a fitted constant. Inset shows the total conductance with $G_{WF}$ subtracted. Above $T\sim80~K$ the conductance departs with a power law of $3.88\pm0.02$. Green and purple dashed lines are the expected thermal conductances due to acoustic phonon (AP) and supercollision (SC) cooling mechanisms, respectively.}
\label{fig:Aria}
\end{figure}

Finally, the capability of our JNT operating on nanoscale devices is demonstrated by measuring the electronic thermal conductance of a graphene device at a wide range of bath temperatures: 3-300~K. Monolayer graphene is mechanically exfoliated, encapsulated in hexagonal boron nitride, and contacted along its 1-dimensional edge~\cite{Dean:2010p2055, *Wang:2013p2560} to form a 2 $\mu$m $\times$ 6 $\mu$m channel. A typical two-terminal channel resistance $R$ of this device varies between 150 and 800 $\Omega$ depending on the back gate voltage. As maximal noise power is collected when the device is impedance matched to the measurement chain, an LC tank circuit is placed on chip to transform the graphene to 50~$\Omega$~\cite{Schoelkopf:1998p2255, Fong:2012p2256} within the measurement bandwidth. The matching network, shown in Fig.~4(a), defines a bandwidth of 25~MHz centered at 133~MHz. For the results shown here, the graphene device is measured away from the charge neutrality point with hole density $n\approx 5.7 \times 10^{10}$cm$^{-2}$, where $R$ varies between 280 and 480~$\Omega$ from 3 to 300~K. The JNT is calibrated to a given sample using the autocorrelation setup shown in Fig.~1(a), following the procedure outlined for the results in Fig.~2. 

Fig.~4(a) shows the electronic thermal conductance measurement schematic. A sinusoidal heating current is injected into the graphene device at 13.7~Hz resulting in a cosinusoidal heating at 27.4~Hz. The thermal conductance between the electronic system and the bath, $G_{th}$, can be found from the relation between the heat flux injected into the graphene via Joule heating, $\dot{Q} =I^2 R$ and the rise in electron temperature, $\Delta T_e$, through Fourier's law: $ \dot{Q} \equiv G_{th} \Delta T_e$. Fig.~4(b) shows the electron temperature as a function of heating current for three different representative bath temperatures: 3~K, 30~K, and 300~K. Solid lines are quadratic fits with the only fitting parameter being $G_{th}$. 

The main panel of Fig.~4(c) shows $G_{th}$ measured for 3~K$<$T$<$300~K.  The measured thermal conductance of our device falls into two temperature regimes. For $T\lesssim100$~K, $G_{th}$ linearly depends on temperature and is well described by the Wiedemann-Franz law. The dashed line in the main panel of Fig.~4(c) shows the theoretical Wiedemann-Franz conductance, $G_{WF}= 12 LT/R$, where $L$ is the Lorenz number and the factor of 12 accounts for the non-uniform temperature rise within the device in this regime~\cite{Yigen:2014p2697, Fong:2013p2542}. We find an experimental Lorenz number of $L =$2.38$\times10^{-8}$W$\Omega$K$^{-2}$, only $3\%$ below the theoretical value $\frac{\pi^2}{3}\left(\frac{k_B}{e}\right)^2$.

At higher temperature, the measured $G_{th}$ becomes substantially larger than $G_{WF}$, indicating a different energy transfer mechanism dominates above 100~K. Fig.~4(c) inset plots $\Delta G\equiv G_{th}-G_{WF}$ versus $T$ in a log-log plot. We observe that $\Delta G\approx G_0 T^\delta$ with an experimentally fitted $\delta = 3.88\pm0.02$ and $G_0=0.23\pm 0.03$~fW/K$^{4.88}$.  We note that the high power law ($\delta\sim$4) in our measured data is in sharp contrast to the sublinear temperature dependence of graphene's lattice conductivity for $T>150$K~\cite{Seol:2010p1987}, suggesting it is unlikely the energy transfer bottleneck and hence the lattice is well thermalized to the bath.
The potential mechanisms of additional heat transport to the thermalized phonons include acoustic phonon (AP) coupling~\cite{Bistritzer:2009p1891, *Viljas:2010p1976}, supercollision cooling (SC)~\cite{Song:2012p2289, *Chen:2012p2291, *Betz:2012p2336, *Graham:2013p2406}, and coupling to optical phonons in the graphene lattice and boron nitride~\cite{Bistritzer:2009p1891, *Viljas:2010p1976, Schiefele:2012p2879}. Given the measured mobility of 60,000~cm$^{2}$V$^{-1}$s$^{-1}$ at 300~K, we estimate the contributions from AP and SC to be an order of magnitude too small to explain the observed $\Delta G$ (see the dashed and dash dot lines in inset of Fig. 4(c)). From our experiment, we estimate the thermal conductivity per unit area at 300~K to be 9.5$\times 10^{4}$ WK$^{-1}$m$^{-2}$. We find this to be comparable to theoretical calculations which suggest that optical phonons, both in the graphene lattice and the boron nitride substrate, may provide an energy relaxation channel substantially larger than acoustic phonons in this low doping, high temperature limit~\cite{Bistritzer:2009p1891, *Viljas:2010p1976, Schiefele:2012p2879}. We also note that previous thermal transport experiments have shown deviations from Wiedemann-Franz conductance in suspended graphene \cite{Yigen:2014p2697} and carbon nanotubes \cite{Santavicca:2010p2360}. 

In summary, we have developed a high frequency, wide bandwidth radiometer capable of both auto- and cross- correlated Johnson noise thermometry at gigahertz frequencies. Using room temperature LNAs and an analog RF linear multiplier, we demonstrated fast and precise temperature measurements reaching a sensitivity of 5.5~mK (0.01\%) in 1~s of integration time. We tested our nanothermometer by measuring thermal transport in hexagonal boron nitride encapsulated, monolayer graphene from 3 to 300~K. Our data suggests strong electron-phonon coupling above 100~K with a high power law close to 4. This high sensitivity noise measurement setup can be also adaptable to study correlation phenomena in mesoscopic devices.

We would like to thank S.~Weinreb for insightful discussion and fabricating the SiGe LNA; B.~Blake and C.~Ryan for digitizer drivers used for comparison to analog measurements; J.~Ravichandran and B.~Kaye for inspiration and critiques. KCF and TAO acknowledge Raytheon BBN Technologies' support on this work. JC acknowledges the support of the FAME Center, sponsored by SRC MARCO and DARPA. XL is supported by DOE (DE-FG02-05ER46215). PK acknowledges a partial support from the Basic Science Research Program (2014R1A1A1004632) through the National Research Foundation of Korea (NRF) funded by the Ministry of Science, ICT and Future Planning.

\end{document}